\documentclass[a4paper,11pt]{article}
\usepackage{jinstpub} 
\usepackage{lineno}
\usepackage{caption}
\usepackage{subcaption}
\usepackage{siunitx}

\title{SiC Based Beam Monitoring System for Particle Rates from kHz to GHz}

\author{Simon Waid, Andreas Gsponer, Jürgen Maier, Philipp Gaggl, Richard Thalmeier, Thomas Bergauer}
\affiliation{Institute of High Energy Physics, Austrian Academy of Sciences, Austria}

\emailAdd{simon.waid@oeaw.ac.at}

\abstract{The extremely low dark current of silicon carbide (SiC) detectors, even after high-fluence irradiation, was utilized to develop a beam monitoring system for a wide range of particle rates, i.e., from the kHz to the GHz regime. The system is completely built from off-the-shelve components and is focused on compactness and simple deployment. Beam tests using a 50 um thick SiC detector reveal, that for low fluences, single particles can be detected and counted. For higher fluences, beam properties were extracted from beam cross sections using a silicon strip detector. Overall accurate results were achieved up to a particle rate of $10^9$ particles per second.}


\notoc
\begin{document}
\maketitle
\flushbottom

\section{Introduction}
The MedAustron accelerator has been designed and built to deliver ion beams for cancer treatment and research \cite{bryantProgressProtonIonMedical1999}. The requirements for the beam differ between research and cancer treatment. Research areas such as microdosimetry or single particle tracking for ion imaging and detector characterization require low-flux beams that allow for the discrimination of single particles impinging onto the detector. Such flow-flux beams have particles rates in the kHz to the MHz regime \cite{ulrich-purCommissioningLowParticle2021}. Patient treatment on the other hand requires significantly higher dose-rates \cite{bryantProgressProtonIonMedical1999}. Currently, proton beam are delivered in  spills with a duration of 10 seconds. During each spills between $0.9 \cdot 10^{10}$ and $1.3 \cdot 10^{10}$ particles are delivered to the treatment room. 

Beam monitors employed in the extraction line at MedAustron are mainly based on scintillating fibers \cite{ulrich-purCommissioningLowParticle2021}. They have proven to be a reliable solution at clinical particle rates. However, they cannot detect low-flux beams used for microdosimetry or detector development \cite{ulrich-purCommissioningLowParticle2021, rossiniCharacterisationScintillatingFibrebased2023}. Thus for commissioning and quality assurance of low-flux beams no beam position monitors are available in the beam line. Commissioning relied solely on beam position monitors installed in the irradiation room. While the commissioning of low flux beams has been accomplished \cite{ulrich-purCommissioningLowParticle2021} the lack of beam monitors in the beam line complicated the commissioning. Quality assurance is hindered by the need for installing a dedicated beam monitor in the irradiation room.

To equip the accelerator with a particle monitor capable of monitoring the low flux beams, 
still being compatible with clinical fluences, one option would be to build a system capable of counting single particles at up to clinical particle rates. Due to fluctuations in the beam intensity such as system would need to be counting single particles at a frequency of up to several \unit{\giga\hertz} over the full detector area. While detector-strips can be made narrow to reduce the particle rate per strip, realistic detectors still require a counting rate of hundreds of \unit{\mega\hertz}. 
An alternative approach is to count single particles only at low fluences. For higher fluences a quantity proportional to the flux, such as a detector current is used. In this work we are presenting such a detector system.

\section{Proposed Beam Monitoring System}
Silicon detectors are finding wide use as single particle detectors. However, when exposed to ionizing radiation, dark current levels in silicon detectors rise rapidly. 
Increasing dark currents are not a problem for single particle detection. The current pulses emitted by the detector due to traversing particles can be easily separated from the dark current by AC coupling. However, measuring the continuous current through the detector due to a high-flux beam impinging onto the detector requires DC coupling. Thus, in our application the dark current cannot be easily separated from the detector current caused by the beam.  This is especially true, as the dark current can be orders of magnitude larger than the current caused by the beam. As a consequence, silicon detectors are not suitable for our application.

Materials not exhibiting the large dark current levels observed for silicon include diamond and silicon carbide \cite{gagglChargeCollectionEfficiency2022}. While diamond is known to be extremely radiation tolerant, it is also extremely expensive. Silicon carbide on the other hand, is now finding widespread use in the semiconductor industry. This has led to high-quality SiC wafers becoming available at sinking cost.  Thus, we chose to base our beam monitor on silicon carbide. 

\begin{figure}
    \centering
     \includegraphics[width=0.7\textwidth]{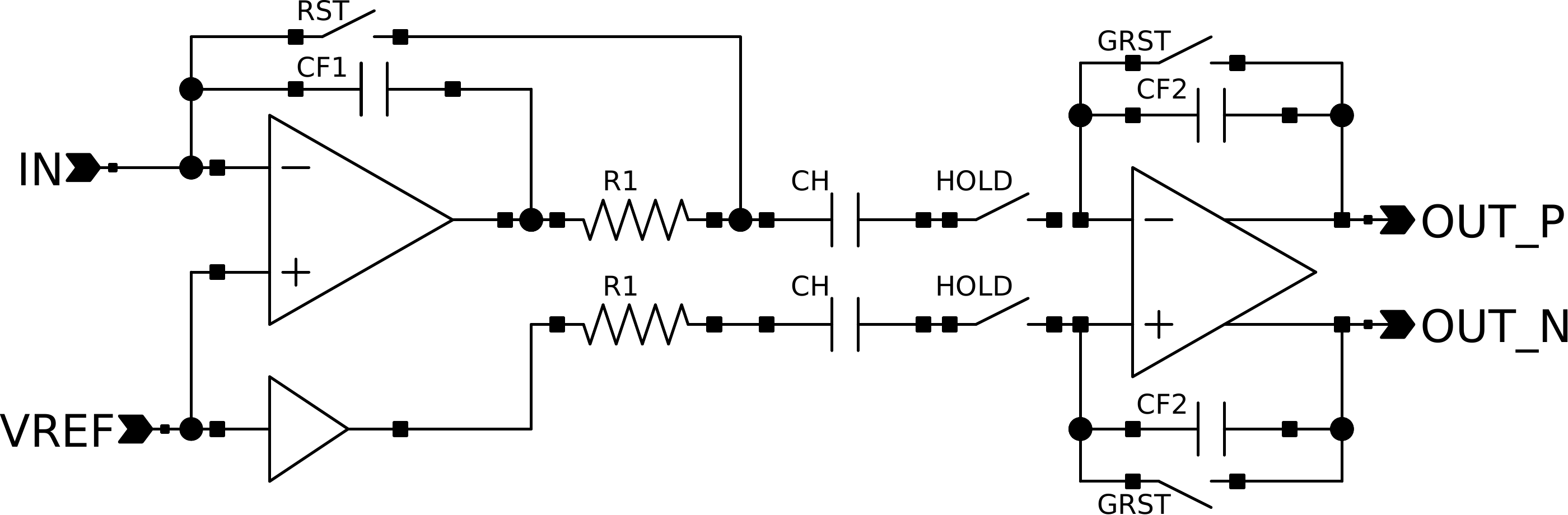}
     \caption{Simplified input circuit of the AD8488 \cite{analogdevicesDataSheetAD84882012}. The input charge is converted to a voltage via CF1. CH and CF2 act as sample/hold element as well as a gain stage. The values of the capacitors CF1 and CH are selectable via configuration registers. This way the amplification of the circuit can be selected in a range from 0.14 to 23 \unit{\volt\per\pico\coulomb}. Similarly, the value of the resistor R1 can be selected in a range from 0 to 195 \unit{\kilo\ohm} resulting in a noise-reducing low-pass filter between R1 and CH \cite{analogdevicesDataSheetAD84882012}. While larger resistor values lead to lower noise, the decreased charging speed of CH leads to a dependence of the output voltage on the time of arrival of particles. }
     \label{fig:ad8488_input}

\end{figure}

 For measuring the current from our detector we chose the Analog Devices  AD8488 X-Ray frontend. This frontend contains 128 low-noise charge integrators and sample/hold elements with selectable gain. A simplified diagram of the AD8488 input circuit is shown in fig. \ref{fig:ad8488_input}. The AD8488 converts an input charge into a voltage (via CF1 in fig. \ref{fig:ad8488_input}). A sample hold stage (CH and CF2 in fig. \ref{fig:ad8488_input}) converts this voltage  back to a charge via CH and back again to a voltage via CF2. This stage provides an output voltage proportional to the input charge. The two-stage conversion circuit enables the amplification to be tuned via two capacitors, CF1 and CH. Both are configurable via registers. This way, the amplification of the circuit can be selected in a range from 0.14 to 23 \unit{\volt\per\pico\coulomb}, corresponding to \qty{44}{\deci\bel} of dynamic range. Further, a low pass for reducing the amplifier noise at the cost of bandwidth can be enabled and configured via registers (R1 in fig. \ref{fig:ad8488_input})  \cite{analogdevicesDataSheetAD84882012}. According to the AD8488 datasheet the root mean square (RMS) noise is below \qty{993}{\elementarycharge}  for a \qty{38}{\pico\farad} thin film transistor (TFT) panel \cite{analogdevicesDataSheetAD84882012}. From SiC we expect \qty{57}{\elementarycharge\per\micro\meter} for a minimum ionizing particle (MIP) \cite{christanell4HsiliconCarbideParticle2022}. The smallest energy deposition in the detector at the Medaustron accelerator is attained when extracting a \qty{800}{\mega\electronvolt} proton beam. The energy deposition under this condition corresponds to 1.2 MIPs. Thus, for \qty{100}{\micro\meter} thick SiC layer, we expect a signal-to-noise ratio (SNR) of 6.9.  Consequently, single particles are expected to be detectable. Given the dynamic range of \qty{44}{\deci\bel}, when reducing the gain of the AD8488, we expect to be able to measure the current through the detector even at clinical fluences.

\begin{figure}
    \centering

    \includegraphics[width=0.7\textwidth]{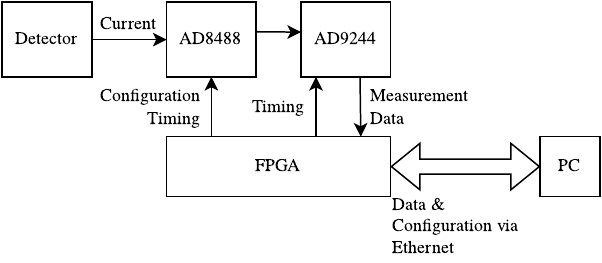}
    \caption{Overall architecture of the system. The detector is read-out using an AD8488 analog frontend. It converts the charges from the detector to voltages. Voltages are converted to digital values using an AD9244 ADC. A Zync Ultrascale+ FPGA controls the timing and configuration of all components, and sends  measurement data to a PC via a conventional gigabit Ethernet interface. 
    }    
    \label{fig:architecture}
\end{figure}

For data acquisition and control we employed a field-programmable gate array (FPGA). We selected a ZU2EG-1E FPGA board obtained from Trenz Electronic GmbH, Germany. An Analog Devices AD9244 14-bit analog-to-digital converter (ADC) was employed to digitize the output signal of the AD8488. A schematic illustration of the data acquisition (DAQ) setup is given in fig. \ref{fig:architecture}. The FPGA board was employed to control the timing and configuration of both the AD8488 and the AD9244 as well as acquiring the measurement values output by the AD9244 ADC. The data is packaged by the FPGA and streamed to a personal computer (PC) using a transfer control protocol (TCP) connection via gigabit Ethernet. The same Ethernet connection was also used for configuring the hardware from the PC via a separate TCP connection. 
Ethernet was chosen, as it allows for easy integration by re-using  existing network infrastructure. To avoid data loss in case of temporary congestion on the Ethernet connection we employed \qty{1}{\giga\byte} of random access memory (RAM) for buffering data prior to transmission. 
The system consists of few components rendering it compact.

Given a shortage in the availability of SiC detectors during our system development phase, tests using SiC detectors were complemented by Si detectors.  A SiC strip detector was used for testing the system in single particle detection mode and for measuring the noise level of the AD8488/AD9244 combination. 
The employed SiC detector had a strip pitch of \qty{100}{\micro\meter} and a strip length of \qty{3}{\milli\meter}. The thickness of the SiC epitaxial layer in which the depletion zone was formed was \qty{50}{\micro\meter}. Thus, at full depletion, the strip capacitance was approximately \qty{0.5}{\pico\farad}. The SiC detector was biased using a Keithley 2470 SMU to attain full depletion. The applied voltage was \qty{400}{\volt}. 

Beam profiles at clinical fluences were measured using an unirradiated, \qty{300}{\micro\meter} thick Si strip detector with a pitch of \qty{100}{\micro\meter} biased at \qty{200}{\volt} to attain full depletion. The dark current induced by radiation damage prior to each spill was extracted from the measurement data and subtracted during post-processing. Only every second strip was connected to the AD8488 to attain a pitch of \qty{200}{\micro\meter}. In future versions of the system, this detector will be replaced by a \qty{100}{\micro\meter} thick SiC detector having a pitch of \qty{250}{\micro\meter}. We assessed, that the Si detector we employed delivered a 1.7 times larger than the signal than we expect from our future SiC detector. 

\section{Results and Discussion}

\begin{figure}
    \centering
    \begin{subfigure}[b]{0.48\textwidth}
         \centering
             \includegraphics[width=\textwidth]{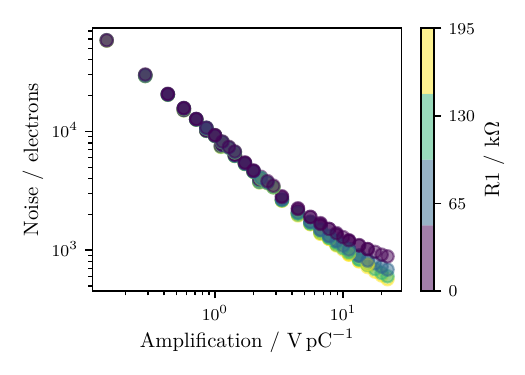}
         \caption{RMS noise in elementary charges.}
         \label{fig:noise_a}
     \end{subfigure}
     \hfill
    \begin{subfigure}[b]{0.48\textwidth}
         \centering
             \includegraphics[width=\textwidth]{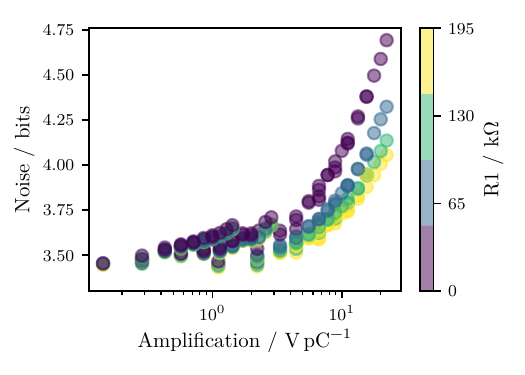}
         \caption{RMS noise in bits.}
         \label{fig:noise_b}
     \end{subfigure}
     \hfill
    
    \caption{Noise of the AD8488 front-end when connected to a small SiC strip detector having a capacitance of \qty{0.5}{\pico\farad}. The noise is given as a function of the selected R1 and the amplification of the AD8488 frontend.}    
    \label{fig:noise}
\end{figure}

We measured the noise levels of the AD8488/AD9244 combination with a SiC strip detector with varying AD8488 settings. The measurement results are given in fig. \ref{fig:noise}. We can see that for the largest gain and R1=\qty{195}{\kilo\ohm} the noise of the AD8488 is approximately \qty{570}{\elementarycharge}. For R1=\qty{0}{\ohm}, the noise increases to \qty{886}{\elementarycharge}. The noise level in ADC counts varies between 3.4 and 4.7 bit. When operating the system in single particle detection mode the noise level in ADC counts is the largest. However, in this operating mode, only the presence or absence of single particles is of interest.
Thus, in terms of ADC bits a high noise level is tolerable. At sensitivities relevant for clinical intensities the SNR in terms of ADC bits is of higher importance. In this measurement range, the noise level is less than 4 bits. The employed ADC has 14 bits overall. One is used for the sign. Thus, the net dynamic range for clinical intensities is better than 9 bits or 55 dB. 

\begin{figure}
    \centering
    \begin{subfigure}[b]{0.48\textwidth}
         \centering
             \includegraphics[width=\textwidth]{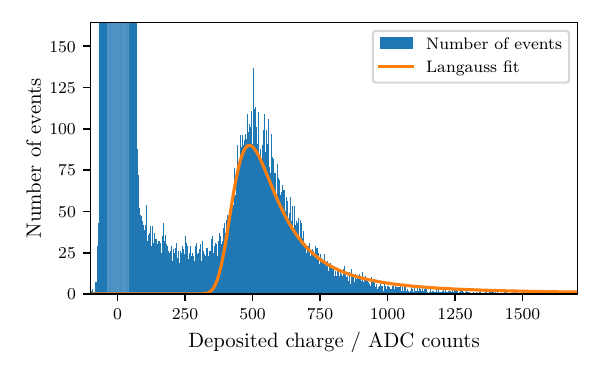}
         \caption{Protons, \qty{62.4}{\mega\electronvolt}, R1=\qty{195}{\kilo\ohm}}
         \label{fig:single_part_a}
     \end{subfigure}
     \hfill
    \begin{subfigure}[b]{0.48\textwidth}
         \centering
             \includegraphics[width=\textwidth]{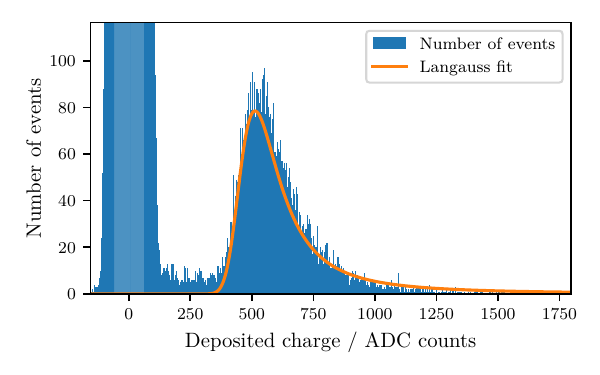}
         \caption{Protons, \qty{62.4}{\mega\electronvolt}, R1=\qty{0}{\ohm}}
         \label{fig:single_part_b}
     \end{subfigure}
     \hfill
    \begin{subfigure}[b]{0.48\textwidth}
         \centering
             \includegraphics[width=\textwidth]{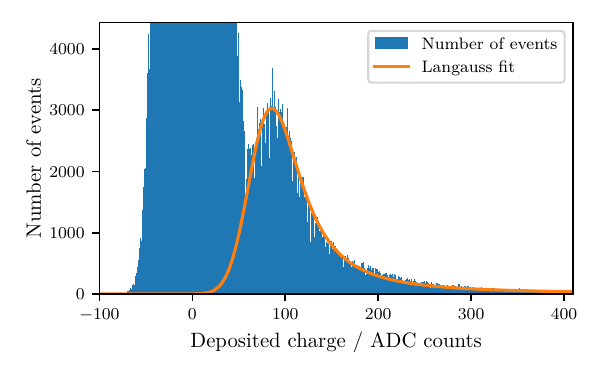}
         \caption{Protons, \qty{800}{\mega\electronvolt}, R1=\qty{195}{\kilo\ohm}}
         \label{fig:single_part_c}
     \end{subfigure}
     \hfill
    \begin{subfigure}[b]{0.48\textwidth}
         \centering
             \includegraphics[width=\textwidth]{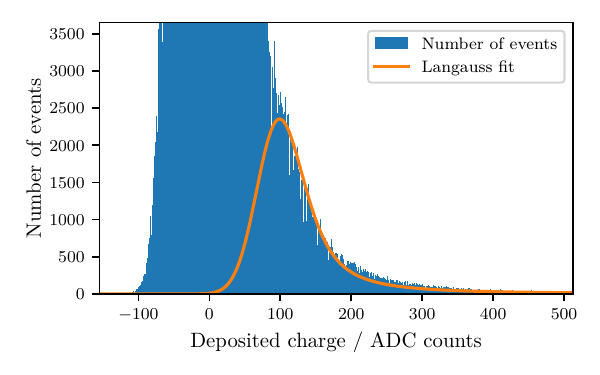}
         \caption{Protons, \qty{800}{\mega\electronvolt}, R1=\qty{0}{\ohm}}
         \label{fig:single_part_d}
     \end{subfigure}
     \hfill

    \caption{Histogram of deposited charge depending on the selected beam energy and the filter setting of the AD8488. For a beam proton beam with an energy of \qty{800}{\mega\electronvolt} and R1=\qty{195}{\kilo\ohm} the SNR was 4.7.  In case of a \qty{62.4}{\mega\electronvolt} choosing R1=\qty{195}{\kilo\ohm} leads to the observation of a deposited charge between the noise peak and the expected langauss distribution. We attribute this to the incomplete charging of the hold capacitor CH when particles hit the detector at the end of the integration interval.}    
    \label{fig:single_part}
\end{figure}

Beam tests in low flux mode were carried out using energies ranging from 62.4 to 800 \unit{\mega\electronvolt}. Histograms of the deposited charge are shown in fig. \ref{fig:single_part}. In case of a \qty{800}{\mega\electronvolt} beam, the SNR was 4.7 when setting R1=\qty{195}{\kilo\ohm}. When replacing the currently employed \qty{50}{\micro\meter} detector with a \qty{100}{\micro\meter} thick device in the future, we expect to improve  this result. With this improvement we expect the SNR to be sufficient for using the detector as a beam monitor. In fig. \ref{fig:single_part_a}  and \ref{fig:single_part_b} show the histogram for a \qty{62.4}{\mega\electronvolt} beam. 
In fig. \ref{fig:single_part_a} R1=\qty{195}{\kilo\ohm}, while in fig. \ref{fig:single_part_b} R1 was set to \qty{0}{\ohm}. One can see that for R1=\qty{0}{\ohm} the peak for particles is clearly delineated from the noise, while for R1=\qty{195}{\kilo\ohm} a continuous spectrum of charge deposition events is visible. We attribute this behaviour to R1 delaying the charging of  the hold capacitor CH. If a particle impinges just before the hold circuit is detached from the input amplifier, the hold capacitor is not yet completely charged.  In case of R1=\qty{0}{\ohm} the AD8488 datasheet indicates a charging time constant of \qty{0}{\micro\second}.
 Consequently, no incomplete charging should take place. This changes with increasing filter resistor R1 to higher values. From this perspective, choosing R1=\qty{0}{\kilo\ohm} would be ideal. In contrast to this at R1=\qty{195}{\kilo\ohm} the noise is lowest. Thus, depending on the particle energy one will have to choose R1 such that a sufficient SNR is attained. For a proton beam having an energy of \qty{800}{\mega\electronvolt} one will have to accept frequent events of incomplete charging while at lower energies one can suppress them by setting R1 to lower values. For a beam monitoring system employed for optics commissioning we deem this behaviour acceptable as information such as beam position and beam shape are unaffected.

\begin{figure}
    \centering
    \begin{subfigure}[b]{0.48\textwidth}
         \centering
             \includegraphics[width=\textwidth]{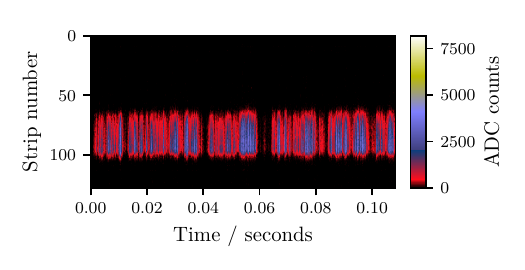}
         \caption{\qty{402}{\mega\electronvolt\per\atomicmassunit} carbon beam}
         \label{fig:x-section_a}
     \end{subfigure}
     \hfill
    \begin{subfigure}[b]{0.48\textwidth}
         \centering
             \includegraphics[width=\textwidth]{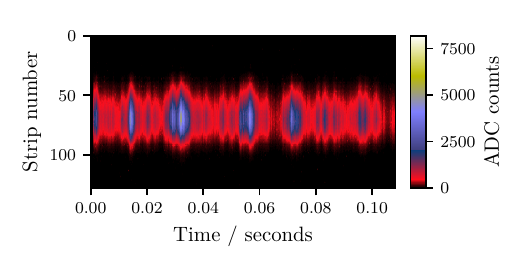}
         \caption{\qty{272.7}{\mega\electronvolt}proton beam}
         \label{fig:x-section_b}
     \end{subfigure}
    
    \caption{Exemplary beam profile in the horizontal plane of clinical beams acquired using the setup. The color shows the recorded intensity.} 
    \label{fig:x-section}
\end{figure}

The system has been tested at clinical rates using proton beams having energies in the range from 62.4 to 252.7 \unit{\mega\electronvolt} and carbon beams with energies in the range of 120 to 402 \unit{\mega\electronvolt}. For this measurements, a Si strip detector was used as no SiC detectors with sufficient size were available. Fig. \ref{fig:x-section}  shows exemplary beam profiles acquired at clinical intensities. 
The dynamic range of the system proved to be sufficient for imaging the beam even at clinical dose rates. 

\begin{figure}
    \centering
    \begin{subfigure}[b]{0.48\textwidth}
         \centering
             \includegraphics[width=\textwidth]{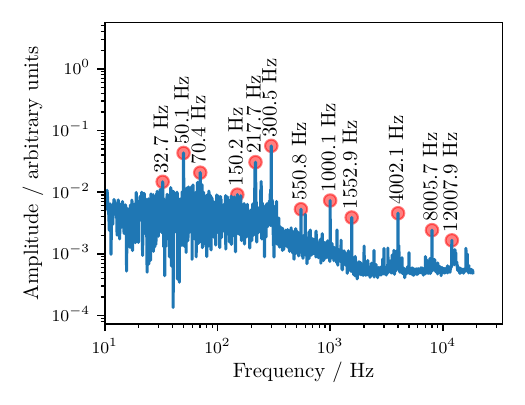}
         \caption{Protons, \qty{119.6}{\mega\electronvolt}, Fourier transform of the COM.}
         \label{fig:analysis_a}
     \end{subfigure}
     \hfill
    \begin{subfigure}[b]{0.48\textwidth}
         \centering
             \includegraphics[width=\textwidth]{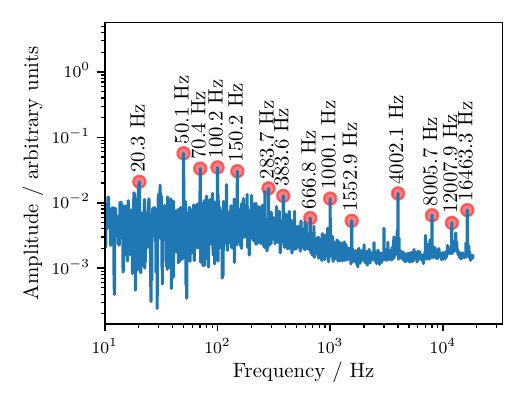}
         \caption{Protons, \qty{119.6}{\mega\electronvolt}, Fourier transform of the FWHM.}
         \label{fig:analysis_b}
     \end{subfigure}
    
    \caption{Exemplary analysis of the acquired beam cross section at clinical fluences. Fourier transforms of fluctuations in beam parameters can be employed for identifying the cause of a fluctuation and mitigating it.}    
    \label{fig:analysis}
\end{figure}

An exemplary analysis on measurement data for extracting  beam parameters relevant for commissioning of accelerator optics was performed. Typical parameters of interest include: (i) intensity profile over time, (ii) center of gravity over time and (iii) beam width over time. Further, performing Fourier transforms can often help in identifying sources of perturbance. Such Fourier transform are exemplary shown in fig. \ref{fig:analysis} for the center of mass (COM) and the full width at half maximum (FWHM) of the beam at an energy of \qty{119.6}{\mega\electronvolt}. In case of the center of mass we can see dominant oscillations at 217.7, 300.5 and 50 \unit{\hertz}. The oscillations at 217.7 and 300.5  \unit{\hertz} can be attributed to dipoles due to their absence in the FWHM spectra. The responsible current sources can be identified by searching for to the oscillations at the mentioned frequencies in the output current of all dipole current sources. Actions can then target specific oscillations on specific current sources.

Our tests using a Si strip detector demonstrated the need for using SiC instead of Si. During out tests, the employed Si strip detector was exposed to approximately 500 spills delivered within 8 hours of beam tests. This radiation exposure led to an increase in dark current equivalent to 50\% of the dynamic range at minimum gain, rendering the detector unusable for most tasks within just a few hours of usage.



\section{Conclusion and Outlook}
We presented a beam monitor made from off-the-shelf components with a high dynamic range. The beam monitor was shown to detect single particles, whereby the SNR for the worst case beam was 4.7 when using a \qty{50}{\micro\meter} thick SiC detector. The beam monitor was further tested at clinical beam intensities where for both carbon and proton beam the spill structure could be resolved. We are currently working on a new version of the system including a \qty{100}{\micro\meter} thick SiC detector. The system is planned to have an active area of 6.4 x 6.4 \unit{\square\centi\meter}{\color{red}.}
\newpage

\acknowledgments
This project has received funding from the Austrian Research Promotion Agency FFG, grant number 883652. Production and development of the 4H-SiC samples was supported by the Spanish State Research Agency (AEI) and the European Regional Development Fund (ERDF), ref. RTC-2017-6369-3.

\bibliographystyle{JHEP}
\bibliography{HEPHY-detector-dev}

\providecommand{\href}[2]{#2}\begingroup\raggedright\begin{thebibliography}{1}

\bibitem{bryantProgressProtonIonMedical1999}
P.J.~Bryant, L.~Badano, M.~Benedikt, P.J.~Bryant, M.~Crescent, P.~Holy et~al.,
  \emph{Progress of the {{Proton-Ion Medical Machine Study}} ({{PIMMS}})},
  \href{https://doi.org/10.1007/BF03038873}{\emph{Strahlentherapie und
  Onkologie} {\bfseries 175} (1999) 1}.

\bibitem{ulrich-purCommissioningLowParticle2021}
F.~{Ulrich-Pur}, L.~Adler, T.~Bergauer, A.~Burker, A.~De~Franco, G.~Guidoboni
  et~al., \emph{Commissioning of low particle flux for proton beams at
  {{MedAustron}}},
  \href{https://doi.org/10.1016/j.nima.2021.165570}{\emph{Nuclear Instruments
  and Methods in Physics Research Section A: Accelerators, Spectrometers,
  Detectors and Associated Equipment} {\bfseries 1010} (2021) 165570}.

\bibitem{rossiniCharacterisationScintillatingFibrebased2023}
R.~Rossini, R.~Benocci, R.~Bertoni, M.~Bonesini, M.~Clemenza, C.~De~Vecchi
  et~al., \emph{Characterisation of a scintillating fibre-based hodoscope
  exposed to the {{CNAO}} low-energy proton beam},
  \href{https://doi.org/10.1016/j.nima.2022.167746}{\emph{Nuclear Instruments
  and Methods in Physics Research Section A: Accelerators, Spectrometers,
  Detectors and Associated Equipment} {\bfseries 1046} (2023) 167746}.

\bibitem{gagglChargeCollectionEfficiency2022}
P.~Gaggl, T.~Bergauer, M.~G{\"o}bel, R.~Thalmeier, M.~Villa and S.~Waid,
  \emph{Charge collection efficiency study on neutron-irradiated planar silicon
  carbide diodes via {{UV-TCT}}},
  \href{https://doi.org/10.1016/j.nima.2022.167218}{\emph{Nuclear Instruments
  and Methods in Physics Research Section A: Accelerators, Spectrometers,
  Detectors and Associated Equipment} {\bfseries 1040} (2022) 167218}.

\bibitem{analogdevicesDataSheetAD84882012}
A.~Devices, \emph{Data {{Sheet AD8488}}},  2012.

\bibitem{christanell4HsiliconCarbideParticle2022}
M.~Christanell, M.~Tomaschek and T.~Bergauer, \emph{{{4H-silicon}} carbide as
  particle detector for high-intensity ion beams},
  \href{https://doi.org/10.1088/1748-0221/17/01/C01060}{\emph{Journal of
  Instrumentation} {\bfseries 17} (2022) C01060}.

\end{thebibliography}\endgroup

\end{document}